# A Parallel-pulling Protocol for Free Energy Evaluation

Van A. Ngo[*]

*University of Southern California, Department of Physics and Astronomy, Los Angeles, California 90089-0242, USA.*

Jarzynski's equality (JE) allows us to compute free energy differences from distributions of work. In molecular dynamics simulations, the traditional way of constructing work distributions is to perform as many pulling simulations as possible. But reliable work distributions are not always produced in a finite number of simulations. The computational cost of using JE is not less than other commonly used methods such as Thermodynamic Integration and Umbrella Sampling methods. Here we first show a different proof of JE based on the idea of step-wise pulling procedures that is efficient in computing free energies by using JE. The key point in our proof is that the processes of turning-on/off a harmonic potential to perform work are described by double Heaviside functions of time. We then show that the distributions of work performed by the potential can be easily generated from the distributions of a reaction coordinate along a pathway. Based on the proof, we propose sequential and parallel step-wise pulling protocols for generating work distributions that require suitable relaxation time at each pulling step. The criterion for reliable work distributions is that there must be sufficient mutual overlaps between the adjacent distributions of the reaction coordinate along the pathway. We arrive at an alternative formula (besides JE) to compute free energy differences from the averaged values of the reaction coordinate. The combination of JE and the alternative formula provides a viable way to determine the accuracy of computed free energy differences. For the stretching of a deca-alanine molecule, our approach requires 21 parallel simulations and relaxation time as small as 0.4 ns for each simulation to estimate free energy differences with an uncertainty of about 13%.

## I. INTRODUCTION

In 1997, Jarzynski [1] showed that the free energy change from an initial configuration **A** to a final configuration **B** can be evaluated from finite-time non-equilibrium measurements by

$$\Delta F_{A \to B} = -\beta^{-1} \ln \langle \exp(-\beta W) \rangle, \quad (1)$$

where $\beta = 1/k_B T$, $T$ is temperature, $k_B$ is Boltzmann constant, $W$ is applied work and $\langle ... \rangle$ denotes the average over all possible trajectories in which work is performed. Those configurations are associated with external control parameters $\lambda$. For simplicity, a system can be characterized by a single control parameter $\lambda$ that is used to monitor pathways of a reaction coordinate. If using a harmonic potential to perform work on the system, parameter $\lambda$ can be the center of the harmonic potential. The center-of-mass position $x$ trapped by the potential in the direction along $\lambda$ is considered as a reaction coordinate. By varying $\lambda$ from $\lambda_1$ (corresponding to configuration **A**) at time $t_1$ to $\lambda_s$ (corresponding to configuration **B**) at time $t_s$, one generates a trajectory in which $W$ is measured from a force-versus-extension curve:

$$W_{\exp} = \int_{\lambda_1}^{\lambda_s} f_\lambda \delta\lambda, \quad (2)$$

where $f_\lambda$ is an applied force measured at a value of $\lambda$ whose increment is $\delta\lambda$. The work definition in Eq. (2) was used to validate Jarzynski's equality (JE) in experiments of stretching a single biological molecule [2-4].

Since the control parameter $\lambda$ is varied, the Hamiltonian $H(\mathbf{z}, \lambda(t))$ of the system accordingly changes with time, where $\mathbf{z} \equiv (\mathbf{q}, \mathbf{p})$ denotes a point in the phase space with **q** and **p** being coordinates and momenta. According to Jarzynski, work $W$ can be evaluated from the Hamiltonian by $W_H = \int_{t_1}^{t_s} [\partial H(\mathbf{z}, \lambda(t))/\partial t] dt$ [1, 3, 5-9]. If one expresses $W$ as $H(\mathbf{z}_s, \lambda_s) - H(\mathbf{z}_1, \lambda_1)$, it is straightforward to prove JE by taking the average of exp($-\beta W$) over the canonical ensemble of $H(\mathbf{z}_1, \lambda_1)$ and applying the canonical transformation for initial $\mathbf{z}_1$ to final $\mathbf{z}_s$. For stochastic processes, $H(\mathbf{z}_s, \lambda_s) - H(\mathbf{z}_1, \lambda_1) = W_{\exp} + Q_{st}$, where the discretized form of $W_{\exp}$ is given by $\sum_{i=1}^{s-1}[H(\lambda_{i+1}, x_i) - H(\lambda_i, x_i)]$ and $Q_{st}$ is heat. To prove JE in this case where $W$ is replaced with the discretized $W_{\exp}$, one must take into account the detailed balance for transition probabilities [7, 10]. To further ascertain the theoretical validity of JE, Crooks [11] provided a general approach, so-called the Crooks fluctuation theorem in which JE can be considered as a special case. The Crooks fluctuation theorem states that the distribution of work $\rho_F(+W)$ in a forward process **A**→**B** and that of work $\rho_R(-W)$ in a reverse process **A**←**B** satisfy the symmetry relation, $\rho_F(+W)/\rho_R(-W) = \exp[\beta(W - \Delta F_{A \to B})]$. Multiplying both sides of the relation with $\rho_R(-W)\exp(-\beta W)$ and integrating over $W$, we obtain Eq. (1). Another approach to prove JE was realized by Hummer and Szabo [12-16]. Their proof is based on the observation of phase space evolution established by Feynman-Kac theorem of statistical mechanics, $\exp[-\beta\Delta F(t_s)] = \langle \exp(-\beta W_H) \rangle$ which is Eq. (1). Hummer and Szabo pointed out that reconstructing free energy profiles from JE must take into account the initial positions of an applied potential. Their conclusion related to the initial potential energy, however, cannot drawn from the former approaches. We report here that there is an alternative approach to prove JE (Section II). We also arrive at the same conclusion as of Hummer and Szabo (Section V).

In order to use Eq. (1) for evaluating the free energy change, one has to construct work distributions from all possible force-versus-extension curves. The remarkable feature of JE is that work distributions can be computed from non-equilibrium processes. Because of this feature, JE provides a powerful tool to compute free energy changes in molecular dynamics simulations in which non-equilibrium processes are often encountered. The main challenge of using JE in molecular dynamics simulations is how to generate reliable work distributions in an efficient way [5, 17-22]. It is computationally expensive to generate all possible trajectories for sampling rare small values of work $W$ that dominate in the average of exp($-\beta W$).

One scheme to overcome the computational difficulty is to implement the Potential of Mean Force (PMF) method developed by Park *et al.* [17, 21] in steered molecular dynamics simulations [23]. In the PMF method $W$ is defined as $-vk\int_0^{t_s}(x-\lambda_1-vt)dt$, where $v$ is a guiding velocity of a harmonic potential with spring constant $k$ and $x$ is a reaction coordinate. In this definition, parameter $\lambda$ is linear with time $t$, i.e., $\lambda \sim vt$. In experiments, this definition of work is valid to estimate the free energy change by Eq. (1). In molecular dynamics simulations, this work is observed to bias free energy changes. The PMF method remedies the biasing problem by utilizing the second order cumulant expansion [17, 21, 24]. The PMF requires that $k$ must be sufficiently large for the expansion. Then the computed free energy differences (FEDs) are unbiased even with a finite number of trajectories.

Though the PMF method has been used, the efficiency and accuracy of computing FEDs in complicated systems remain unsatisfactory. The efficiency is defined as computational cost of evaluating FEDs for given error tolerance. Rodriguez-Gomez and Darve [18, 25] showed in the simulations of transfer of fluoromethane across a water-hexane interface that the accuracy of the PMF method is poor in comparison with their adaptive biasing force method. Bastug *et al.* [26, 27] showed that FEDs computed from a finite number of pulling trajectories are biased. The cost of evaluating the FEDs using the PMF method is higher than that of using Umbrella Sampling with Weighted Histogram Analysis method [28-30]. Oberhofer *et al.* [19] analyzed that the efficiency of simulations based on JE cannot compete with the ones of the Umbrella Sampling and Thermodynamic Integration [24, 31] methods.

Recently, Hernandez and colleagues [32] developed an adaptive steered molecular dynamics for improving the accuracy of the PMF method. In this adaptive scheme, instead of applying JE for single pulling trajectories they applied JE in series of shorter steps of the pulling trajectories. At each $c$-th step, a system is allowed to relax for a certain amount of time before performing next pulling step. The configuration for next pulling step is chosen so that it minimizes the work difference, $\left|\Delta F^c - W_\alpha^c(t)\right|$, where $\Delta F^c$ and $W_\alpha^c(t)$ are free energy change and work done at $c$-th step in $\alpha$-th trajectory at time $t$, respectively. This scheme essentially suggests that rare small work distributions can be generated by breaking a long pathway into shorter ones and letting systems relax to minimize the biasing effects of an applied potential on free energy changes. Similar to the original PMF method, this scheme fails if work distributions are not trivial Gaussian functions; hence the second-order cumulant expansion cannot be used.

In this article we propose a different approach to prove JE and to compute free energy changes more efficiently than the PMF methods. We take the definition of $W_H$ as a starting point and treat control parameter $\lambda$ and time $t$ in different manners. We introduce double Heaviside functions of time $t$, divide a pathway into a series of steps and take into account relaxation time at each step. The Heaviside functions are used to describe the procedures of turning-on/off a harmonic potential to perform work. A series of steps and relaxation time are crucial to generate rare small values of work. The approach results in an identity $\langle\exp(-\beta W_H)\rangle = 1$ [Eq. (10)] that helps to prove JE, $\Delta F_{A\rightarrow B} = -\beta^{-1}\ln\langle\exp(-\beta W_{exp})\rangle$ [Eq. (12)]. Our theory also suggests an approximation to compute FEDs (besides JE) from the averaged values of a reaction coordinate [Eq. (14)]. This approximation resembles the Thermodynamic Integration.

Based on the theory, we propose sequential and parallel step-wise pulling protocols to reconstruct work distributions. We show that it is straightforward to produce work distributions from the distributions of a reaction coordinate without any adaptive scheme. Based on the simulation results in a test case, we provide a criterion that requires sufficient overlaps between the adjacent distributions of a reaction coordinate for ensuring reliable work distributions.

In comparison with the original and adaptive PMF methods, our scheme of generating work distributions and computing FEDs does not require large values of $k$ and the assumption of Gaussian distributions of work. Given that relaxation time is long enough (several ns) and the number of discretized pulling steps is adequate, FEDs can be accurately computed. Our simulation results show that the parallel-pulling protocol speeds up free energy calculations with an acceptable accuracy.

Our article is organized as follows: we present the theory in Section II; Section III highlights two methods based on the theory to compute FEDs and describes sequential and parallel pulling protocols; the theory is tested in the simulations of unfolding a deca-alanine molecule in Section IV; Section V provides discussions on the theory and methods.

## II. THEORY

Let us consider a system of $N$ particles described by a time-independent Hamiltonian $H_o(\vec{p}^{3N},\vec{r}^{3N})$, where $\vec{p}^{3N}$ and $\vec{r}^{3N}$ are the $3N$ dimensional momenta and coordinates, respectively. The system has a canonical ensemble with temperature $T$. An external harmonic potential $U(x,\lambda_1) = k(x - \lambda_1)^2/2$ can be applied to a set of trapped particles, whose center-of-mass position along $x$-direction is defined as a reaction coordinate $x$. Here $\lambda_1$ is the center of the harmonic potential (control parameter) and $k$ is the spring constant. We suppose that there is no coupling between the harmonic potential with the particles' momenta. Then our partition functions $Z(\lambda_i,k)$ ($i \geq 1$) can be simply expressed in terms of spatial coordinates $(\vec{r}^{3N-1},x)$ for calculating free energy differences (FEDs).

From time $t_0$ to $t_1$, we apply the harmonic potential to the system. One can in principle take the infinity limit for $t_0$ and $t_1$. As a result, the coupling Hamiltonian has the following formula:

$$H(\lambda_1,x) = H_o(\vec{r}^{3N-1},x) + \frac{k}{2}(x-\lambda_1)^2 \theta(t-t_0)\theta(t_1-t), \quad (3)$$

where $\theta(t)$ is a Heaviside step function. Here we have omitted momenta in $H_o$ for simplicity. Then the work applied to the system from a specific state at $t = t_0$ to a final state is computed as

$$W_H = \int_{t_0}^{t_1} \frac{\partial H(\lambda_1,x)}{\partial t} dt = \frac{k}{2}[(x_0-\lambda_1)^2 - (x_1-\lambda_1)^2], \quad (4)$$

where $x_0$ is an initial value at time $t_0$ of the reaction coordinate belonging to the ensemble with $H_o(\vec{r}^{3N-1},x_0)$ and $x_1$ is any final value of the reaction coordinate at $t_1 \geq t_0$ belonging to the ensemble with $H(\lambda_1,x_1)$. The applied force exerted on the reaction coordinate along $x$-direction at time $t_1$ is $f(x_1,\lambda_1) = -\partial U(x_1,\lambda_1)/\partial x_1$.

Given that the canonical ensemble of $x_1$ is generated after turning on the harmonic potential, one can take the average of $\exp(-\beta W_H)$ ($\beta = 1/k_B T$) over the canonical ensemble (see Appendix A) to obtain

$$\left\langle \exp[\beta \frac{k}{2}(x_1-\lambda_1)^2] \right\rangle_{(x_1,\lambda_1,k)} = \exp[\beta \Delta F(\lambda_1,k)], \quad (5)$$

where $(x_1,\lambda_1,k)$ represents all possible points with Hamiltonian $H(\lambda_1,x_1)$ in phase space and $\Delta F(\lambda_{i=1},k) = F(\lambda_{i=1},k) - F_0 = -\beta^{-1}\ln[Z(\lambda_{i=1},k)] + \beta^{-1}\ln[Z(0)]$. The partition functions are $Z(\lambda_i,k) = \int d\vec{r}^{3N-1} dx_i \exp[-\beta H(\lambda_i,x_i)]$ and $Z(0) = \int d\vec{r}^{3N-1} dx_0 \exp[-\beta H_o(\vec{r}^{3N-1},x_0)]$. $\Delta F(\lambda_{i=1},k)$ is the FED between the configurations with and without the harmonic potential. Alternatively, given the canonical ensemble of $x_0$ the average can be taken over all $x_0$ instead of $x_1$ to evaluate the same FED:

$$\Delta F(\lambda_1,k) = -\beta^{-1} \ln \left\langle \exp[-\beta \frac{k}{2}(x_0-\lambda_1)^2] \right\rangle_{(x_0,k=0)}, \quad (6)$$

where $(x_0,k=0)$ represents all possible points with Hamiltonian $H_o(\vec{r}^{3N-1},x_0)$ in phase space. From Eqs. (5) and (6) along with inequality $\langle \exp(-\beta W_H)\rangle \geq \exp(-\beta \langle W_H\rangle)$ [33], we have the lower and upper bounds to $\Delta F(\lambda_1,k)$:

$$\left\langle \frac{k}{2}(x_1-\lambda_1)^2 \right\rangle_{(x_1,\lambda_1,k)} \leq \Delta F(\lambda_1,k)$$
$$\leq \left\langle \frac{k}{2}(x_0-\lambda_1)^2 \right\rangle_{(x_0,k=0)}. \quad (7)$$

Now, we perform a series of steps with $\lambda_i$ ($i = 1, 2...s$) to pull the reaction coordinate by moving the center of the harmonic potential from $\lambda_1$ to $\lambda_s$. At each step with $\lambda_i$ the system is relaxed from time $t_{i-1}$ to $t_i$. Then the center is instantaneously shifted from $\lambda_i$ to $\lambda_{i+1}$ at time $t_i$. As a result, the total work (see Appendix B) is

$$W_{total} = \sum_{i=1}^{s} \frac{k}{2}\left[(x_{i-1}-\lambda_i)^2 - (x_i-\lambda_i)^2\right]$$
$$= \frac{k}{2}\left[(x_o-\lambda_1)^2 - (x_s-\lambda_s)^2\right] + W_{mech}, \quad (8)$$

where mechanical work $W_{mech}$ in each generated trajectory is defined by

$$W_{mech} = \frac{k}{2}\sum_{i=1}^{s-1}(\lambda_{i+1}-\lambda_i)(\lambda_{i+1}+\lambda_i - 2x_i) \approx \int_{\lambda_1}^{\lambda_s} f_\lambda \delta\lambda, \quad (9)$$

where $f_\lambda = \partial U(x,\lambda)/\partial \lambda$ for a sufficiently small increment $\delta\lambda$. When $s$ is equal to 1, $W_{mech}$ is zero and Eqs. (5) or (6) should be used to estimate FEDs. Equation (9) can be expressed as $\sum_{i=1}^{s-1}[H(\lambda_{i+1},x_i) - H(\lambda_i,x_i)]$ that was used to prove Jarzynski's equality (JE) in stochastic processes [7, 10].

It should be noted that the canonical ensembles with $H(\lambda_i,x_i)$ for different $\lambda_i$ are independent of one another. Hence, one can take average of $\exp(-\beta W_{total})$ over all the canonical ensembles of $x_0, x_1...x_s$ at the same time to arrive at the following identity:

$$\langle \exp(-\beta W_{total})\rangle_{(x_0,x_1,...,x_s)} = 1. \quad (10)$$

Alternatively we first average $\exp(-\beta W_{total})$ over the ensemble of $x_0$ with $H_o(\vec{r}^{3N-1},x_0)$ and the ensemble of $x_s$ with $H(\lambda_s,x_s) = H_o(\vec{r}^{3N-1},x_s) + U(x_s,\lambda_s)$ to obtain

$$\langle \exp(-\beta W_{total})\rangle_{x_0,x_s} = \exp\{\beta[F(\lambda_s,k) - F(\lambda_1,k)]\}$$
$$\times \exp(-\beta W_{mech}). \quad (11)$$

We then average the left hand side of Eq. (11) over all the rest of $x_i$ and make use of the equality Eq. (10) to arrive at

$$\langle \exp(-\beta W_{mech})\rangle_{FVE} = \exp[-\beta \Delta F^{JE}(\lambda_s,\lambda_1,k)], \quad (12)$$

where $\Delta F^{JE}(\lambda_s,\lambda_1,k) = F(\lambda_s,k) - F(\lambda_1,k)$ with $F(\lambda_i,k) = -(1/\beta)\ln[Z(\lambda_i,k)]$ and $\langle ...\rangle_{FVE}$ is the average over all values of $W_{mech}$ measured from force-versus-extension (FVE) curves (detailed derivation of Eq. (12) is given in Appendix B). Equation (12) is identical to JE, Eq. (1).

It is noted that the distributions of $x_i$ have the property of randomness due to thermal fluctuations. According to the probability theory of independent random numbers [34], $W_{mech}$'s distribution $\rho(W_{mech})$ is proportional to $\rho(x_1)\rho(x_2)...\rho(x_{s-1})$ where $\rho(x_i)$ are $x_i$'s distributions. The relation between $\rho(W_{mech})$ and $\rho(x_i)$ means that $\rho(W_{mech})$ can be constructed if a set of $\rho(x_i)$ is known.

Since the reaction coordinate is trapped by the potential, $\rho(x_i)$ can be approximated as $\exp[-\beta k(x_i - \langle x_i \rangle)^2/(2\gamma_i^2)]$, where $\langle x_i \rangle$ is the averaged position of the reaction coordinate at $i^{th}$ pulling step and $\gamma_i^2$ is $k\sigma_i^2/k_B T$ with $\sigma_i$ equal to the standard deviation of $x_i$'s distribution. Thus $\rho(W_{mech})$ can be expressed in terms of $x_i$, $\langle x_i \rangle$ and $\gamma_i^2$ with $i$ from 1 to $s-1$. Given the distribution $\rho(W_{mech})$ and based on Eq. (12) we derive a simple relationship between Gaussian-approximated FED $\Delta F^G(\lambda_s, \lambda_1, k)$ and $\langle x_i \rangle$ (see Appendix B):

$$\Delta F^G(\lambda_s, \lambda_1, k) = \frac{k(\lambda_s - \lambda_1)^2}{2s} \frac{\sum_{i=1}^{s-1}(1-\gamma_i^2)}{s}$$
$$+ k(\lambda_s - \lambda_1) \frac{\sum_{i=1}^{s-1}(\lambda_i - \langle x_i \rangle)}{s}. \quad (13)$$

The first term in Eq. (13) vanishes as $s$ goes to infinity, i.e., infinitely slow pulling limit. Then the FED $\Delta F^G(\lambda_s, \lambda_1, k)$ can be determined from the second term,

$$\Delta F_{fluct}(\lambda_s, \lambda_1, k) = k(\lambda_s - \lambda_1) \frac{\sum_{i=1}^{s-1}(\lambda_i - \langle x_i \rangle)}{s}, \quad (14)$$

which only depends on the average of the differences between $\lambda_i$ and $\langle x_i \rangle$. In that limit, the right hand side of Eq. (14) becomes the Thermodynamic Integration $\int_{\lambda_1}^{\lambda_s} \langle \partial H(\lambda, x)/\partial \lambda \rangle_\lambda d\lambda$. This limit implies that the required relaxation time $\tau_i = t_i - t_{i-1}$ at each $\lambda_i$ can be arbitrarily small. Moreover, if all $\gamma_i^2$ are unity, $\Delta F^G(\lambda_s, \lambda_1, k)$ can also be determined from $\Delta F_{fluct}(\lambda_s, \lambda_1, k)$ even with a finite value of $s$. $\gamma_i^2$ equal to unity indicates that $x_i$'s distributions resemble the canonical distributions of the system ($\sim \exp[-\beta H(\lambda_i, x_i)]$). The condition for $\gamma_i^2$ equal to unity is satisfied when $\tau_i$ is large (the assumption for canonical distributions). These two limiting cases of $\tau_i$ suggest that the first term in Eq. (13) is not important and Eq. (14) can be used to evaluate FEDs with finite $s$ and finite $\tau_i$.

### III. Protocols

Based on the theory in the previous section, we propose two methods for free energy calculations: (i) from the distributions of mechanical work $W_{mech}$ that is defined by Eq. (9) used for JE, Eq. (12); (ii) from averaged values of $\langle x_i \rangle$ used for Eq. (14).

According to JE, or Eq. (12), it is necessary to generate as many pulling trajectories as possible to construct work distributions. As illustrated by Fig. 1a, a conventional protocol is that $N_p$ pulling trajectories from configuration **A** to configuration **B** are performed in parallel. Unfortunately, there has been no criterion to determine a minimum of $N_p$. Equation (5) supports that a free energy difference (FED) $\Delta F(\lambda_i, k)$ can be calculated when a complete canonical ensemble of $x_i$ is sampled regardless of pathways from $x_o$ [35]. But using Eq. (5) requires large relaxation time $\tau_i$.

Alternatively, according to the expression of $W_{mech}$ [Eq. (9)] work distributions can be produced from the distributions of $x_i$. This expression implies that rare small values of work are not available if the increments $\lambda_{i+1} - \lambda_i$ are larger than the magnitude of $x_i$'s fluctuations, which is defined as the distance from the center of $x_i$'s distribution to its right end. Therefore, the adjacent distributions of $x_i$ must mutually overlap to generate rare small values of work.

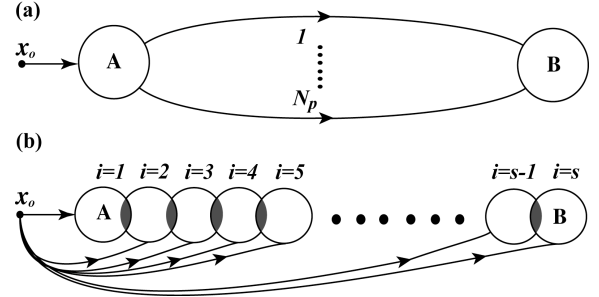

FIG. 1. Schematic illustration for (a) original parallel pulling (arrowed arcs) protocol (b) our proposed sequential (series of circles) and parallel pulling (arrowed arcs) protocols. A reaction coordinate, $x_o$, represents any state without virtual harmonic potentials. The arcs with arrows represent parallel pulling trajectories. **A** and **B** are initial and final configurations, respectively. $N_p$ is a number of trajectories. The circles represent sets of states or distributions of the reaction coordinate. Shaded areas are mutual overlapping states. Index $i$ denotes the ordered number of pulling and runs from 1 to $s$, where indices 1 and $s$ also represent the initial and final configurations, respectively.

Guided by the argument we propose two protocols (illustrated in Fig. 1b) to construct reliable work distributions with a small number of pulling trajectories and small values of $\tau_i$:

**(i)** Sequential pulling protocol (SPP): given a pathway that can be characterized by a single control parameter $\lambda$ (the center of a harmonic potential) in a system, we assign $\lambda_1$ to the configuration **A**, $\lambda_s$ to the configuration **B**. We then divide the pathway into $s-2$ intermediate configurations (steps) that are characterized by different $\lambda_i$. We let the harmonic potential pull the reaction coordinate by sequentially assigning $\lambda$ to be $\lambda_i$ with $i$ from 1 to $s$ in a single simulation. At each step corresponding to $\lambda_i$, the system is relaxed for time $\tau_i$ to collect distributions of $x_i$. All possible values of $x_i$ are used to generate work distributions whose algorithm is given in Appendix C.

**(ii)** Parallel pulling protocol (PPP): given a pathway and a set of different $\lambda_i$ corresponding to $s$ targeted configurations as in **(i)**, $s$ pulling simulations are independently carried out to pull the reaction coordinate from a state $x_o$ to the targeted configurations. In each simulation, the center of a harmonic potential $\lambda$ is independently assigned to be $\lambda_i$. Each simulation is run for adequate relaxation time $\tau_i$ to collect distributions of $x_i$ corresponding to $\lambda_i$. The same procedure to construct work distributions from $x_i$ is given in Appendix C.

Once work distributions are constructed, it is straightforward to evaluate FEDs based on JE. By measuring all averaged values of $x_i$ corresponding to $\lambda_i$, FEDs are also estimated by Eq. (14).

## IV. IMPLEMENTATION AND TESTING

In this section we test the two protocols on an exemplary system: helix-coil transition of deca-alanine in vacuum at $T = 300$ K (see Ref. [17] for more details of simulation setups). NAMD2 package [36] and CHARMM22 force fields [37] are used here. We generate work distributions and compute free energy profiles by Eqs. (12) and (14) in the sequential and parallel pulling protocols. We also investigate the effects of spring constants $k$ and relaxation time $\tau_i = \tau$ for all pulling steps.

### A. Sequential pulling protocol (SPP)

The simulation setups are the same as in Ref. [17]. One end of the molecule is kept fixed at the origin. The other end is sequentially pulled by a harmonic potential having the spring constant $k = 7.2$ kcal/mol/Å$^2$ (we use this potential through the paper otherwise mentioned). The position along the pulling direction of the pulled end is considered as a reaction coordinate. We increase $\lambda$ by $\Delta\lambda = 0.5$ Å from $\lambda_1 = 13$ to $\lambda_s = 33$ Å ($s = 41$). At each step with $\lambda_i$ ($i = 1, 2…s$) the system is relaxed for $\tau = 10$ ns. We record the position of the pulled end $x_i$ every 0.1 ps at each relaxation step. Some of distributions $x_i$ are shown in Fig. 2a. Figure 2b shows the initial ($\lambda = 13$ Å), transition ($\lambda = 25$ Å) and final ($\lambda = 33$ Å) configurations. We use Least Square Fitting method to obtain $\gamma_i^2 = k\sigma_i^2/k_BT$ as shown in Fig. 2c, where $\sigma_i$ is equal to the standard deviation of $x_i$'s distributions and $k_B$ is Boltzmann constant. The dimensionless values of $\gamma_i^2$ fall in the range of 1.1 to 1.4 and are peaked at $\lambda = 26.5$ Å.

We divide the ranges of $x_i$ and $W_{mech}$ into small bins to construct work distributions (see Appendix C). The range for $x_i$ is from 0 to 50.0 Å with a bin width $\delta x = 0.001$ Å. The range for $W_{mech}$ is from -150 to 150.0 kcal/mol with a bin width $\delta W = 0.01$ kcal/mol. The work distributions change as $\lambda$ increases (Fig. 3a). Plugging these distributions and measured $\langle x_i \rangle$ into Eqs. (9), (12), (13) and (14), we compute applied mechanical work $W_{mech}$, $\Delta F^{JE}$, $\Delta F^G$ and $\Delta F_{fluct}$ as functions of $\lambda$, respectively. The referenced free energy for those free energy differences (FEDs) is $F(\lambda_1,k)$ that is used through out the paper. $\lambda$ and $k$ are omitted in the functions for a simple notation. The Potential of Mean Force (PMF) method [17] is used to generate the exact PMF profile with the pulling speed of $v = 0.1$ Å/ns.

Figure 3b shows that the profiles of $\Delta F^{JE}$ and $\Delta F_{fluct}$ are in a good agreement with the PMF profile. At the minimum position ($\lambda \sim 15$ Å) $\Delta F^{JE}$ is the same as that of the PMF profile (~ –2.1 kcal/mol) that is smaller than $\Delta F_{fluct}$ by 0.5 kcal/mol. At $\lambda \geq 25$ Å (kink position), the values of the PMF profile are larger than $\Delta F_{fluct}$ by 1.0 kcal/mol, but less than $\Delta F^{JE}$ ($\geq 16$ kcal/mol) by 0.5 kcal/mol. Free energies $\Delta F^G$ are smaller than the others because the effects of finite $s$ and $\tau$ cause a significant contribution to $\Delta F^G$. The smaller values of $\Delta F^G$ and the agreement between the profile of $\Delta F_{fluct}$ and the PMF profile confirm the argument at the end of Section II that the first term in Eq. (13) is negligible.

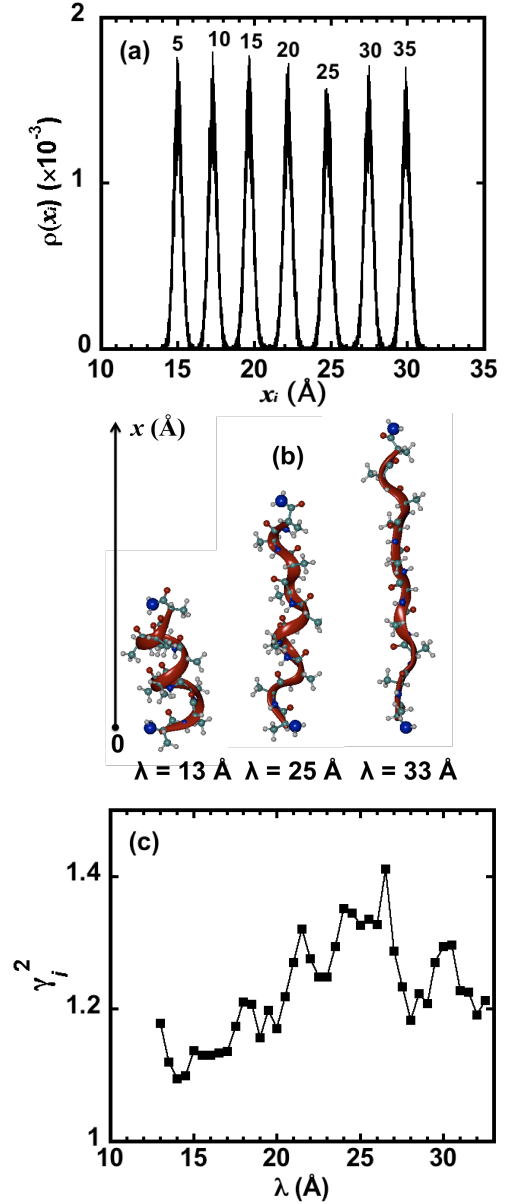

FIG. 2. (a) Normalized distribution of $x_i$. The numbers indicate the pulling steps $i$. (b) Configurations of deca-alanine at $\lambda = 13$, 25 and 33 Å. The two biggest balls are fixed and pulled ends. (c) Dimensionless quantity $\gamma_i^2$ versus $\lambda$.

The variance of $\Delta F^{JE}$ can be estimated by $\sigma_W^2/Q + \beta^2 \sigma_W^4/2(Q-1)$ [18, 38], where $\sigma_W$ is the standard deviations of work distributions and $Q$ is a number of bins that have non-zero values. Then the standard deviations of $\Delta F^{JE}$ are about 6% of $\Delta F^{JE}$. But the uncertainty of $\Delta F_{\text{fluct}}$ cannot be computed as $(k_B T/k)^{1/2} \sum_{i=1}^{s-1} \gamma_i / |\sum_{i=1}^{s-1} (\lambda_i - \langle x_i \rangle)|$ that are unreasonably large (> 100%). Since $\Delta F_{\text{fluct}}$ is an approximation from $\Delta F^{JE}$, its uncertainty should be comparable to that of $\Delta F^{JE}$. The uncertainty of $\Delta F^{JE}$ can be reduced when $Q$ is larger. But we observe that the difference between $\Delta F^{JE}$ and $\Delta F_{\text{fluct}}$ is not changed at larger values of $Q$. This suggests that the difference can be chosen as an uncertainty of the FEDs ($\Delta F_{\text{fluct}}$ and $\Delta F^{JE}$) if it is larger than that of $\Delta F^{JE}$.

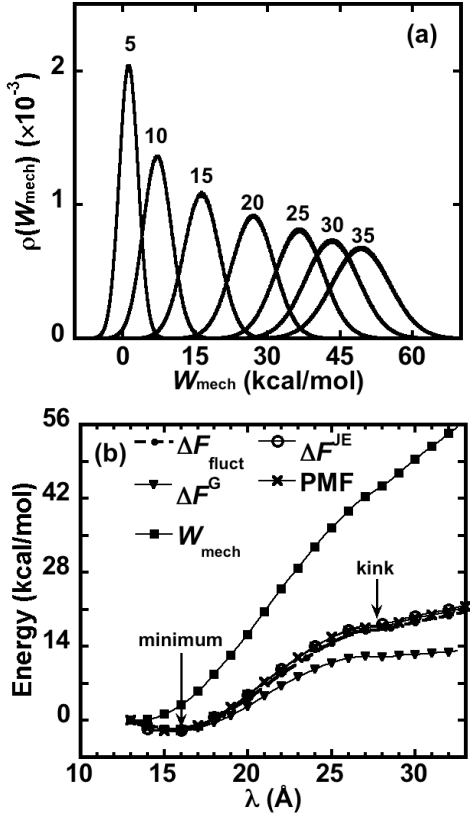

FIG. 3. (a) Normalized work distribution. The numbers indicate the pulling steps $i$. (b) Mechanical work (squares) and free energies versus $\lambda$. PMF (crosses), $\Delta F^{JE}$ (empty circles), $\Delta F^G$ (triangles) and $\Delta F_{\text{fluct}}$ (dots) are free energies computed by the PMF method, Eq. (12), Eq. (13) and Eq. (14), respectively.

It is noted that the highest value of $\gamma_i^2$ occurs at the same position where the free energy profiles are observed to be a kink ($\lambda \sim 26.5$ Å). The variation of $\gamma_i^2$ indicates how the approximated widths of $x_i$'s distributions relatively change along a pathway. Therefore, any changes of $\gamma_i^2$ can be used to examine how the overlapping portions of $x_i$'s distributions vary.

### B. Parallel pulling protocol (PPP)

The agreement among the FEDs computed by the exact PMF method, $\Delta F_{\text{fluct}}$ [Eq. (14)] and $\Delta F^{JE}$ [Eq. (12)] suggests that one can compute FEDs if the values of $\langle x_i \rangle$ and their distributions are available in any pulling protocols. To verify this observation, we perform 21 10ns-simulations in parallel of stretching the molecule from the same initial state $x_o$. In the simulations 21 values of $\lambda_i$ are assigned to be 13, 14…33 Å ($\Delta\lambda = 1$ Å and $s = 21$). Each simulation characterized by a single value of $\lambda_i$ is independent of the others. In each simulation we record values of $x_i$ in every 0.1 ps for constructing work distributions. We use $\tau = 10$ ns.

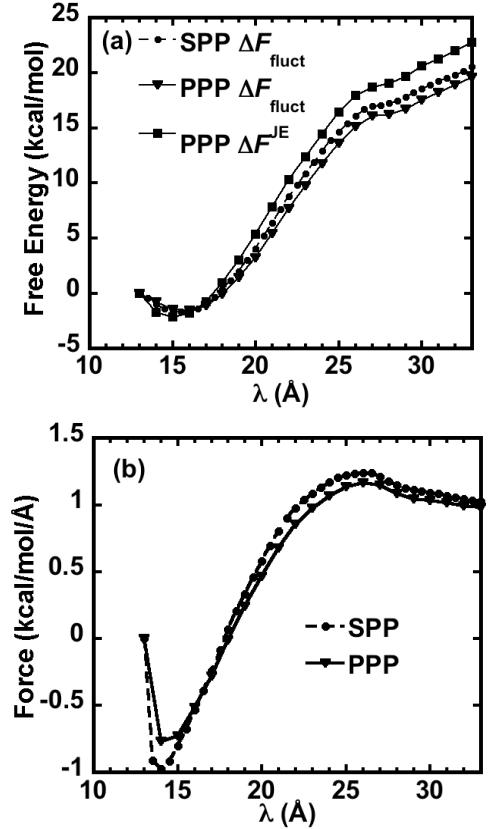

FIG. 4. (a) Free energies versus $\lambda$. $\Delta F_{\text{fluct}}$ (dots) are in the SPP with $\Delta\lambda = 0.5$ Å. $\Delta F_{\text{fluct}}$ (triangles) and $\Delta F^{JE}$ (squares) are in the PPP with $\Delta\lambda = 1$ Å. (b) Accumulating forces $\Delta F_{\text{fluct}}/(\lambda_s - \lambda_1) = k \sum_{i=1}^{s-1} (\lambda_i - \langle x_i \rangle)/s$ versus $\lambda$ in both protocols. The force at $s = 1$ is set to zero.

The computed profile of $\Delta F_{\text{fluct}}$ (triangles) is plotted in Fig. 4a together with $\Delta F_{\text{fluct}}$ (dots) obtained in Section IV A. Figure 4a shows that $\Delta F_{\text{fluct}}$ in both SPP and PPP have the same minimum FED. For $\lambda > 15$ Å, the profile of $\Delta F_{\text{fluct}}$ in the PPP is gradually shifted below the one in the SPP by an amount of 1.0 kcal/mol. With the data of $x_i$ we generate work distributions and evaluate $\Delta F^{JE}$ as also plotted in Fig. 4a. The minimum value of $\Delta F^{JE}$ is slightly smaller than those of $\Delta F_{fluct}$. The values of $\Delta F^{JE}$ in the PPP are 2.0 kcal/mol larger than those of $\Delta F_{fluct}$ in the SPP at $\lambda \geq 25$ Å. The uncertainty of $\Delta F^{JE}$ is about 6%.

We observe that the separation between the profiles of $\Delta F^{JE}$ and $\Delta F_{fluct}$ becomes smaller as $\Delta \lambda$ is changed from 1.0 Å (see Fig. 4a) to 0.5 Å (see Fig. 3b). The more accurate profile of $\Delta F_{fluct}$ in the SPP lies between the profiles of $\Delta F^{JE}$ and $\Delta F_{fluct}$ in the PPP. This suggests that the $\Delta F^{JE}$ and $\Delta F_{fluct}$ in the PPP can be combined to estimate an accurate free energy profile of the system. The free energies of the combined profile can be defined as $(\Delta F^{JE}+\Delta F_{fluct})/2$. Subsequently, the uncertainty of these free energies would be equal to $(\Delta F^{JE} - \Delta F_{fluct})/2$ that is at most 1.5 kcal/mol (~ 10%) as $\Delta\lambda = 1.0$ Å. In terms of computational cost, the PPP is two times faster and more efficient than the SPP given large computational resources.

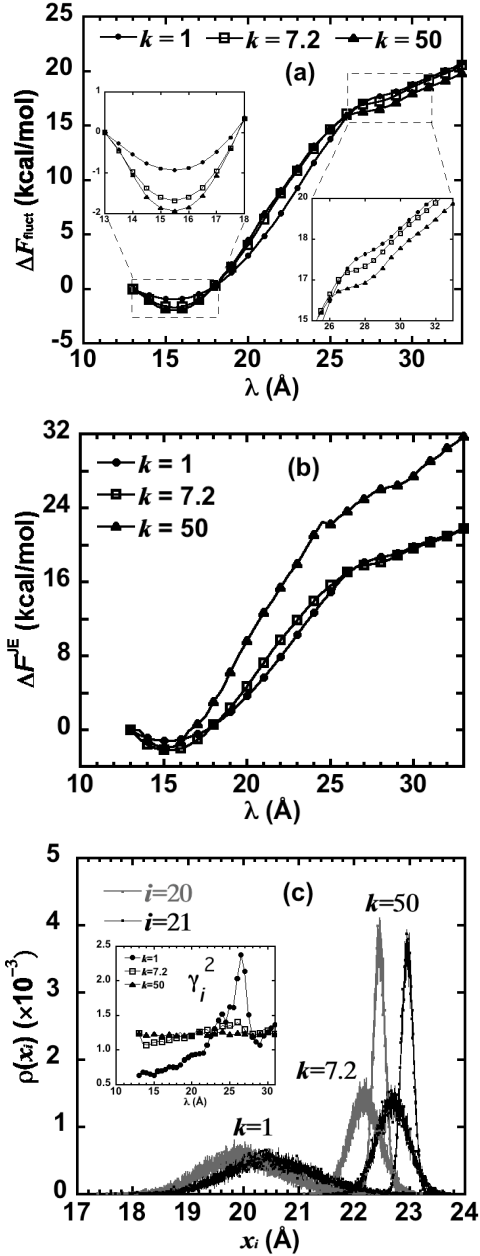

FIG. 5. Free energy profiles of (a) $\Delta F_{fluct}$ and (b) $\Delta F^{JE}$ with different spring constants $k=1.0$ (dots), 7.2 (empty boxes) and 50.0 (triangles) kcal/mol/Å$^2$. (c) Normalized distributions of $x_i$ at step $i = 20$ (gray) and 21 (black) in the three cases. The insets in (a) are zoomed at the minimum and kink positions. The inset in (c) shows $\gamma_i^2$ versus $\lambda$.

It is worth comparing the averaged applied forces in both SPP and PPP. The factor $k\sum_{i=1}^{s-1}(\lambda_i - \langle x_i \rangle)/s$ in Eq. (14) can be interpreted as averaged accumulating forces. Figure 4b shows that the accumulating forces in both the pulling protocols are almost the same.

In order to examine the effect of the magnitude of increment $\Delta\lambda$ on the free energies, we collect 11 distributions of $x_i$ with $\Delta\lambda = 2$ Å ($s = 11$, $\lambda_i = 13, 15…33$ Å) out of the 21 distributions to construct corresponding work distributions. As a result, the FEDs at $\lambda = 25$ Å computed from this data set using JE and Eq. (14) are about 31.8 and 11.4 kcal/mol, respectively. With these data we overestimate the FED if using JE and underestimate it if using Eq. (14). These FEDs clearly indicate that the data set with $\Delta\lambda = 2$ Å cannot be used to construct correct distributions of work and measure a sufficient history of $\langle x_i \rangle$. This observation is consistent with the implication of the expression of $W_{mech}$ that the increments $\Delta\lambda = \lambda_i - \lambda_{i-1}$ should not exceed the magnitude of $x_i$'s fluctuations, which is defined as the distance from the center of $x_i$'s distribution to its right end; otherwise rare small values of $W_{mech}$ are not available. Figure 2a shows that the magnitude of $x_i$'s fluctuations is about 1.0 Å. Consequently $\Delta\lambda$ should not be greater than 1.0 Å. It means that $s$ can be finite but not arbitrarily small.

### C. Effects of spring constant $k$

In order to investigate how FEDs vary in response to different spring constants $k$, we perform two sequential pulling simulations in which $k = 1$ and 50 kcal/mol/Å$^2$, $\Delta\lambda = 0.5$ Å and $\tau = 10$ ns [39].

We observe that the sets of $\langle x_i \rangle$ in all the three cases are not the same. In the case of $k = 1$ kcal/mol/Å$^2$ $\langle x_i \rangle$ are not as perfectly linear with $\lambda$ as in the other cases. But the free energy profiles of $\Delta F_{fluct}$ as plotted in Fig. 5a are in a good agreement with one another. The minimum FEDs are –0.9, –1.6 and –1.9 kcal/mol for $k = 1$, 7.2 and 50 kcal/mol/Å$^2$, respectively. At $\lambda \geq 25$ Å, the three curves converge and shift by less than 1.0 kcal/mol from each other with the same order as at the minimum position (see the insets in Fig. 5a). The smallest spring constant results in the higher values of the FEDs at the minimum and kink positions ($\lambda \sim 15$ and 26 Å), while the strongest one gives rise to the smaller values. With the uncertainty of 1.0 kcal/mol, we confirm that those important values evaluated by Eq. (14)

are independent of spring constants in the range from 1 to 50 kcal/mol/Å$^2$, even though there is noticeable lowering of the FEDs at $k = 1$ kcal/mol/Å$^2$ along the pathway from the minimum to the kink positions.

As computed by JE [40], the free energy profiles of $\Delta F^{JE}$ with $k = 1$ and 7.2 kcal/mol/Å$^2$ agree with each other, whereas the profile with $k = 50$ kcal/mol/Å$^2$ is noticeably higher than the others (see Fig. 5b). To explain the significant distinction among the profiles, we look at the distributions of $x_i$ at step $i = 20$ and 21 ($\lambda \sim 23$ Å) where the transition is about to occur. Figure 5c shows that the two $x_i$'s distributions at $k = 1$ kcal/mol/Å$^2$ have the largest mutual overlapping, whereas at $k = 50$ kcal/mol/Å$^2$ they have much less mutual overlapping. The values of $\gamma_i^2$ in the inset of Fig. 5c indicate how $x_i$'s distributions relatively change along the pathway. The smaller values of $\gamma_i^2/k$ indicate narrower widths of $x_i$'s distributions since the same increment $\Delta\lambda = 0.5$ Å is used. The properties of those distributions suggest an explanation for the distinction based on the expression of $W_{mech}$ in Eq. (9). With large spring constants, the harmonic potential confines the reaction coordinate in such narrow regions that its distributions have little mutual overlapping.

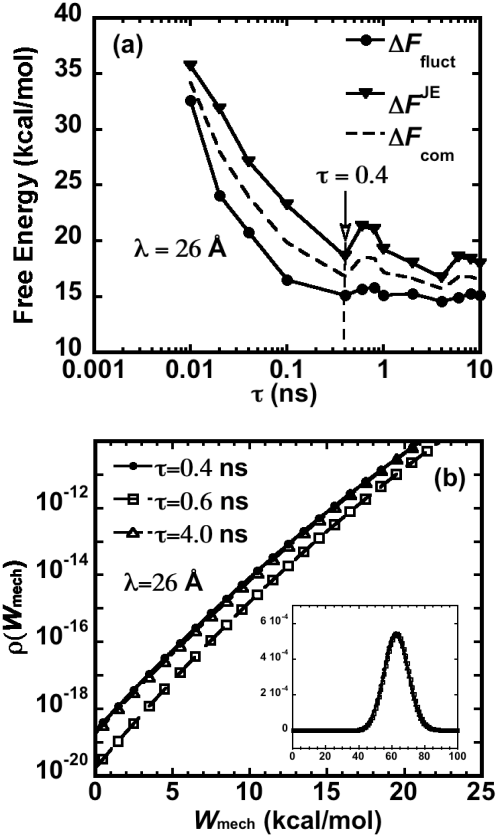

FIG. 6. (a) Free energies $\Delta F_{fluct}$ (dots), $\Delta F^{JE}$ (full triangles) and $\Delta F_{com} = (\Delta F_{fluct} + \Delta F^{JE})/2$ (dashed line) at $\lambda = 26$ Å versus relaxation time $\tau$. (b) Normalized work distributions at $\lambda = 26$ Å in a range from 0 to 25 kcal/mol at $\tau = 0.4$ (dots), 0.6 (empty boxes) and 4.0 (empty triangles) ns. The inset in (b) shows the overall work distributions at the three values of $\tau$.

Consequently, large $k$ reduces the number of rare small values of work, thus increases the magnitude of computed FEDs. We notice that the values of $\Delta F^{JE}$ can also noticeably increase if $\tau$ is too small. Therefore it is essential to investigate the effects of $\tau$ on the FEDs.

### D. Effects of relaxation time $\tau$

We repeat the PPP in Section IV B by reducing relaxation time $\tau$ from 10 ns to 0.01 ns to optimize required simulation time for the test case. Spring constant $k = 7.2$ kcal/mol/Å$^2$ and increment $\Delta\lambda = 1.0$ Å ($s = 21$) are used in the simulations. We evaluate FEDs $\Delta F_{fluct}$ and $\Delta F^{JE}$ at $\lambda = 26$ Å (kink position). Here we consider those FEDs as functions of $\tau$.

Figure 6a shows how the FEDs change in response to variation of $\tau$. The curve of $\Delta F^{JE}$ is noticeably higher than the curve of $\Delta F_{fluct}$ that starts converging at $\tau = 0.4$ ns. The convergence of $\Delta F_{fluct}$ is due to the strong applied harmonic potential that makes the corresponding sets of $\langle x_i \rangle$ not essentially change even at $\tau = 0.4$ ns. But $\Delta F^{JE}$ does not converge as smoothly as $\Delta F_{fluct}$ does. The values of $\Delta F^{JE}$ have small peaks at $\tau = 0.6$ and 6.0 ns. These peaks indicate some important changes in the work distributions. For example, in Fig. 6b the work distribution at $\tau = 0.6$ ns in a range from 0 to 25 kcal/mol is smaller than those at $\tau = 0.4$ and 4.0 ns. These values of work have major contributions to $\Delta F^{JE}$. The work having smaller distributions gives rise to larger $\Delta F^{JE}$. Nevertheless, the inset of Fig. 6b shows no visible deviation among their overall distributions.

In addition, we recall that the separation between the profiles of $\Delta F^{JE}$ and $\Delta F_{fluct}$ reduces as we change $\Delta\lambda$ from 1.0 Å to 0.5 Å (see Figs. 3b and 4a). It means that $\Delta F^{JE}$ and $\Delta F_{fluct}$ give us an upper bound and a lower bound to the FED, respectively. This suggests that an accurate value of the FED at $\lambda = 26$ Å would be somewhere between the $\Delta F^{JE}$ and $\Delta F_{fluct}$ as shown by the dashed line in Fig. 6a. We define a combined FED as $\Delta F_{com} = (\Delta F^{JE} + \Delta F_{fluct})/2$. This combined FED starts converging at $\tau = 0.4$ ns. Therefore, the uncertainty of $\Delta F_{com}$ is equal to $(\Delta F^{JE} - \Delta F_{fluct})/2$ or $\sim 2/16 \sim 13\%$ at $\tau = 0.4$ ns, which is larger than that of $\Delta F^{JE}$ ($\sim 8\%$).

### V. DISCUSSIONS AND CONCLUSIONS

The theory contains a different treatment of the correlation between control parameter $\lambda$ and time $t$. We suggest that $\lambda$ in general should rather be a time-independent parameter characterizing perturbed states. Finite relaxation time $\tau$ is required to let a system evolve into those perturbed states. To account for the overall change of parameter $\lambda$ with time, one can use double Heaviside functions of time $t$. The introduction of the double Heaviside functions describes the physics of how gradually a system absorbs $W_H$ from an applied harmonic potential over time (see Section II). The use of a single Heaviside function of time does not correctly reflect the process of performing work by using harmonic potentials [41-43].

It should be clearly noted here that our definition of work by Eq. (9) agrees with the integral expression of work, $-vk\int_0^{t_s}(x-\lambda_1-vt)dt$ or in general $\int_{t_1}^{t_s}[\partial H(\mathbf{z},\lambda)/\partial\lambda](d\lambda/dt)dt$ where $v$ is a guiding velocity, $k$ is spring constant, $\lambda_1$ is an initial fixed value of $\lambda$, $x$ is a reaction coordinate, $H(\mathbf{z},\lambda)$ is a total Hamiltonian with $\mathbf{z}$ being a point in the phase space, and $\lambda$ is here a function of time. The discretized expression of the work in Eq. (9) is the same as the one used in stochastic processes [7, 10]. Using the discretized expression we proved JE without adopting the concept of detailed balance for transition probabilities. For a large number of pulling steps, $\lambda$ can be expressed as $\lambda_1 + vt$ in Eq. (9) to recover the integral expression of work used in the steered molecular dynamics simulations [17, 21, 32].

The resulting relationship between Hamiltonian-related work $W_{total} = W_H = \int_{t_1}^{t_s}[\partial H(\mathbf{z},\lambda(t))/\partial t]dt$ and $W_{mech}$ is given by Eq. (8). The theory suggests that $W_{total} = W_H \neq \int_{t_1}^{t_s}[\partial H(\mathbf{z},\lambda)/\partial\lambda](d\lambda/dt)dt = W \equiv W_{mech}$. As indicated by the identity, Eq. (10), $W_{total}$ should not appear in Eq. (1) as $W$. This identity suggests that one can add any heat bath effects [44] to systems, which do not influence applied potentials, and normalize the effects to unity. Then the free energy changes can be always computed from the distributions of $W_{mech}$ in non-equilibrium processes regardless of heat bath effects. Jarzynski and Crooks using different approaches arrived at the same conclusion [1, 5-11, 43]. The identity appears to not agree with the expression derived by Hummer and Szabo [15] using Feynman-Kac theorem, $\exp[-\beta\Delta F(t_s)] = \langle\exp(-\beta\int_0^{t_s}[\partial H(\mathbf{z},t)/\partial t]dt)\rangle$ where $\langle\ldots\rangle$ denotes the average over all possible trajectories. The difference stems from our introduction of the double Heaviside functions to the coupling Hamiltonian $H(\mathbf{z},t)$ (Section II).

It is noted that whenever an applied harmonic potential is turned on, the free energy of a system always increases by $\Delta F(\lambda_1,k)$ as indicated by the inequalities (7). We define $G(\lambda_s,\lambda_1) = \Delta F^{JE}(\lambda_s,\lambda_1,k) - \Delta F(\lambda_1,k)$ as a state free energy function. From Eqs. (5) and (12), $G(\lambda_s,\lambda_1)$ can be written as

$$G(\lambda_s,\lambda_1) = -\beta^{-1}\ln\{\langle\exp[-\beta(\int_{x_0}^{x_1}fdx - U(x_0,\lambda_1))]\rangle_{(x_1,\lambda_1,k)} \times \langle\exp(-\beta\int_{\lambda_1}^{\lambda_s}fd\lambda)\rangle_{FVE}\}, \quad (15)$$

where $f$ is $-\partial U(x,\lambda_1)/\partial x$, $x_0$ is an initial position of the reaction coordinate and $x_1$ is a position of the reaction coordinate at the first pulling step. Equation (15) looks very similar to the most important result derived by Hummber and Szabo, i.e., the expression of Eq. (45) in Ref. [15]. The first average is equivalent to the average over all first-time-turning-on-potential procedures. Nevertheless, there is a slight difference between the second averages of Eqs. (15) and (45). In the average $\langle\ldots\rangle_{FVE}$ the mechanical work is an integral taken over $\lambda$ instead of $x$ for all force-versus-extension (FVE) curves (as noted by Jarzynski [9]). One more note here is that the Eq. (15) is exact whereas Eq. (45) is an approximation. To some extent, the similarity between the two equations suggests that Eq. (15) can be used to reconstruct free energy landscapes in the same way as proposed by Hummer and Szabo. It means that the initial positions of an applied potential must be taken into account to reconstruct free energy landscapes.

We have shown that it is possible to construct work distributions from the distributions of a reaction coordinate (Sections III and IV). An acceptable agreement of $\Delta F^{JE}(\lambda_s,\lambda_1,k)$ with the exact PMF profile (Figs. 3b and 4a) confirms the validity of the sequential and parallel pulling protocols. In the protocols, one first has to choose an ordered set of targeted configurations that can be characterized by different values of $\lambda$. One then independently or sequentially turns on the harmonic potential to drive the system into those targeted configurations from any chosen state $x_o$. The protocols do not require any additional task in choosing a typical structure for each pulling step as implemented in the adaptive steered molecular dynamics simulations [32]. All distributions of a reaction coordinate are equally treated to reconstruct work distributions. A number $s$ of pulling steps and an amount of relaxation time $\tau$ are decisive to generate rare small work distributions.

In terms of computational cost, the parallel pulling simulations (Section IV Part B) are two times faster than the sequential ones (Section IV Part A) and have the computational cost equal to performing 100 trajectories with pulling speed $v = 10$ Å/ns using the PMF method [17].

In the PMF method, a finite number ($N_p$) of parallel forward-pulling trajectories, as illustrated in Fig. 1a, would produce free energy profiles that are likely dependent of pulling speeds [17, 18, 26, 27]. It has been elusive to guarantee that the pulling protocol produces rare states that have a major contribution to work distributions (or what should be a minimum of $N_p$?). If work distributions are not Gaussian functions, the PMF method fails. Kofke *et al*. [45, 46] suggested that the overlap of the work distributions constructed in forward- and reverse-pulling procedures should be adequate to ensure the convergence of exponential-work averages in Eq. (1). In other words, one might have to perform simulations of reverse pulling (folding) for examining the accuracy of work distributions constructed in forward pulling (un-folding) simulations. It is possible to carry out a few reverse-pulling trajectories in experiments and use the Crooks fluctuation theorem to examine work distributions [3, 11, 22, 47]. But in simulations, such reverse-pulling trajectories can be expensive.

Based on the theory and the simulation results in Section IV, we propose a criterion that the overlaps between successive $x_i$'s distributions should be appropriately comparable to their standard deviations, as schematically described in Fig. 1b and plotted in Fig. 5c. The more the adjacent distributions overlap, the more rare small values of work are available. If the sequential and parallel step-wise pulling protocols meet the criterion, one can carry out only forward-pulling simulations to produce reliable work distributions. These work distributions do not depend on pulling speeds $v$.

The theory also suggests that one can combine trajectories with different pulling speeds, which are generated by the PMF method. Because the data [$x(t_1)$ and $x(t_2)$] of any two trajectories with $v_1$ and $v_2$ belong to the same control parameter $\lambda_1 + v_1 t_1$ (or the same perturbed state), where $v_1 t_1 = v_2 t_2$. These data play the same role in the proposed protocols.

In addition, we have shown that averaged values of a reaction coordinate can be used for estimating free energy differences (FEDs) based on Eq. (14). In the limit of slow pulling, i.e., infinitely small $\Delta\lambda = (\lambda_s - \lambda_1)/s$, Eq. (14) reduces to the well-known Thermodynamic Integration [24, 31, 48, 49] and is similar to the Adaptive Biasing Force (ABF) equations [18, 25]:

$$\Delta F_{\text{fluct}}(\lambda_s, \lambda_1, k) \xrightarrow{\Delta\lambda \to 0} \int_{\lambda_1}^{\lambda_s} \left\langle \frac{\partial H(\lambda, x)}{\partial \lambda} \right\rangle_\lambda d\lambda$$

$$= -\int_{\lambda_1}^{\lambda_s} \langle f_a(x) \rangle_\lambda d\lambda = \Delta F_{\text{ABF}}, \quad (16)$$

where $-\langle f_a(x) \rangle_\lambda$ are averaged adaptive forces exerted on a reaction coordinate and $\Delta F_{\text{ABF}}$ is the FED computed by the ABF method. In the ABF method one has to collect all possible external forces from $s$ windows in comparison with only $s-1$ ones as in Eq. (14). Hence, the difference between $\Delta F_{\text{fluct}}(\lambda_s, \lambda_1, k)$ and $\Delta F_{\text{ABF}}$ at the same finite increment $\Delta\lambda$ is $\Delta\lambda \langle f_s \rangle \approx -\Delta\lambda \langle f_a(x) \rangle_{\lambda_s}$ where $\langle f_s \rangle$ [in Eq. (9)] is the final averaged force measured at $\lambda = \lambda_s$. The error of the ABF method is proportional to $\Delta\lambda \sigma_{f_a} \sqrt{s(1+2\kappa)\delta t/\tau}$, where $\sigma_{f_a}$ is the standard deviation of $f_a(x)$, $\kappa$ is the correlation length [50] of the forces and $\delta t$ is a time-step for collecting forces. Accordingly, with a sufficiently large number of available data, the error of $\Delta F_{\text{fluct}}(\lambda_s, \lambda_1, k)$ should be proportional to $\Delta\lambda \left( |\langle f_s \rangle| + \sigma_{f_\lambda} \sqrt{s(1+2\kappa)\delta t/\tau} \right)$, where $\sigma_{f_\lambda}$ is the standard deviation of $f_\lambda$.

The ABF method does not require specifying any forms of applied forces that are adaptive or constrained along pathways. In our method [Eq. (14)], no constraint is imposed on applied forces. Importantly, the behaviors of the averaged accumulating forces (Fig. 4b) in the sequential and parallel pulling protocols are almost the same. Moreover, $\tau$ can be small (~ns) to collect a set of $\langle x_i \rangle$ for free energy calculations (Fig. 6a). These two results suggest that it is possible to estimate FEDs by performing parallel-pulling protocols to measure $\langle x_i \rangle$ or averaged accumulating forces without imposing any constraints on external forces.

The introduction of Eq. (14) for estimating FEDs turns out to be useful to examine the reliability of work distributions used in JE, Eq. (12). As shown in Section IV, the accuracy of FEDs depends upon a number $s$ of pulling steps, relaxation time $\tau$ and spring constant $k$. A viable method to estimate the accuracy is to compare FEDs evaluated by both JE and Eq. (14). Given finite $s$ and $\tau$, JE might overestimate FEDs and Eq. (14) underestimates FEDs (Figs. 4 and 6). Equation (14) is used in the test case for an acceptable accuracy at many values of spring constant $k$ (Fig. 5a) and relaxation time $\tau$ as small as 0.4 ns. FEDs computed by JE are unbiased when $k$ is small and $\tau$ is large (Figs. 5b and 6a). Using both equations, one can estimate FEDs that are equal to $\Delta F_{\text{com}}(\lambda_s, \lambda_1, k) = [\Delta F^{\text{JE}}(\lambda_s, \lambda_1, k) + \Delta F_{\text{fluct}}(\lambda_s, \lambda_1, k)]/2$ with an uncertainty equal to $[\Delta F^{\text{JE}}(\lambda_s, \lambda_1, k) - \Delta F_{\text{fluct}}(\lambda_s, \lambda_1, k)]/2$. By checking the convergence of $\Delta F_{\text{com}}(\lambda_s, \lambda_1, k)$ with respect to $\tau$, one can confirm the accuracy of computed FEDs (Fig. 6a).

In conclusion, we provided an alternative proof of Jarzynski's equality (JE). The key point in our proof is that the processes of turning-on/off a harmonic potential to perform work are described by double Heaviside functions of time. The important results of the proof [Eqs. (9), (12) and (15)] are consistent with the established theories on JE. Our theory also suggests a formula [Eq. (14)] to evaluate free energy differences from averaged values of a reaction coordinate. Our contributions to computational studies of JE are:

**(i)** Work distributions are simply constructed from the distributions of a reaction coordinate.
**(ii)** We proposed sequential and parallel step-wise pulling protocols to generate work distributions.
**(iii)** To use Eqs. (12) (JE) and (14) with a finite number of pulling steps and reasonably small relaxation time, the mutual overlapping range between the adjacent distributions of a reaction coordinate must be comparable to their standard deviations. The more the adjacent distributions overlap, the more rare small values of work are available.
**(iv)** The combination of Eqs. (12) and (14) can be used to estimate free energy changes with an uncertainty equal to half the difference between them.
**(v)** We showed in a test case that our method requires 21 parallel simulations and relaxation time as small as 0.4 ns for each simulation to estimate free energy profiles with an uncertainty of about 13%.

**ACKNOWLEDGMENTS**

We especially acknowledge Dr. Nakano for valuable discussions and great help in organizing the paper. We thank Drs. Choubey, Vemparala, Nomura and Shing for helpful suggestions.

## APPENDIX A: SINGLE PULLING STEP

Here we derive Eqs. (5) and (6) using canonical ensembles of $x_0$ and $x_1$. Averaging of $\exp(-\beta W_H)$ ($\beta = 1/k_B T$) over $x_1$'s ensemble is computed as

$$\left\langle e^{-\beta W_H} \right\rangle_{(x_1, \lambda_1, k)} = \frac{\int d\vec{r}^{3N-1} dx_1 e^{-\beta\left[W_H + H_o(\vec{r}^{3N-1}, x_1) + \frac{k}{2}(x_1 - \lambda_1)^2\right]}}{\int d\vec{r}^{3N-1} dx_1 e^{-\beta\left[H_o(\vec{r}^{3N-1}, x_1) + \frac{k}{2}(x_1 - \lambda_1)^2\right]}}$$

$$= \frac{\int d\vec{r}^{3N-1} dx_1 e^{-\beta\left[\frac{k}{2}(x_0 - \lambda_1)^2 + H_o(\vec{r}^{3N-1}, x_1)\right]}}{Z(\lambda_1, k)}$$

$$= e^{-\beta \frac{k}{2}(x_0 - \lambda_1)^2} \frac{\int d\vec{r}^{3N-1} dx_1 e^{-\beta H_o(\vec{r}^{3N-1}, x_1)}}{Z(\lambda_1, k)},$$

which is equivalent to

$$\left\langle \exp\left\{-\beta \frac{k}{2}\left[(x_0 - \lambda_1)^2 - (x_1 - \lambda_1)^2\right]\right\}\right\rangle_{(x_1, \lambda_1, k)}$$

$$= \exp\left[-\beta \frac{k}{2}(x_0 - \lambda_1)^2\right] \exp[\beta \Delta F(\lambda_1, k)].$$

Since a single value of $x_0$ at time $t_0$ is constant in the average, we cancel both sides the factors related to $x_0$. Thus, we arrive at Eq. (5):

$$\left\langle \exp\left[\beta \frac{k}{2}(x_1 - \lambda_1)^2\right]\right\rangle_{(x_1, \lambda_1, k)} = \exp[\beta \Delta F(\lambda_1, k)].$$

A similar procedure can be carried out to derive Eq. (6):

$$\left\langle e^{-\beta W_H} \right\rangle_{(x_0, k=0)} = \frac{\int d\vec{r}^{3N-1} dx_0 e^{-\beta\left[W_H + H_o(\vec{r}^{3N-1}, x_0)\right]}}{\int d\vec{r}^{3N-1} dx_0 e^{-\beta H_o(\vec{r}^{3N-1}, x_0)}}$$

$$= e^{\beta \frac{k}{2}(x_1 - \lambda_1)^2} \frac{\int d\vec{r}^{3N-1} dx_0 e^{-\beta\left[\frac{k}{2}(x_0 - \lambda_1)^2 + H_o(\vec{r}^{3N-1}, x_0)\right]}}{Z(0)},$$

which is equivalent to

$$\Delta F(\lambda_1, k) = -\beta^{-1} \ln \left\langle \exp\left[-\beta \frac{k}{2}(x_0 - \lambda_1)^2\right]\right\rangle_{(x_0, k=0)}.$$

## APPENDIX B: SERIES OF PULLING STEPS

The total work absorbed by the system in a series of pulling steps is given by

$$W_{\text{total}} = \int_{t_0}^{t_s} \frac{\partial H(\lambda_1, \lambda_2 ... \lambda_s, x)}{\partial t} dt = \int_{t_0}^{t_s} \frac{\partial}{\partial t}[H_o(\vec{r}^{3N-1}, x)$$

$$+ \frac{k}{2} \sum_{i=1}^{s}(x - \lambda_i)^2 \theta(t - t_{i-1})\theta(t_i - t)] dt$$

$$= \frac{k}{2}\left[(x_0 - \lambda_1)^2 - (x_s - \lambda_s)^2\right]$$

$$+ \frac{k}{2} \sum_{i=1}^{s-1}\left[(\lambda_{i+1} - x_i)^2 - (x_i - \lambda_i)^2\right]$$

$$= \frac{k}{2}\left[(x_0 - \lambda_1)^2 - (x_s - \lambda_s)^2\right]$$

$$+ \sum_{i=1}^{s-1}\left[H(\lambda_{i+1}, x_i) - H(\lambda_i, x_i)\right]. \quad (B.1)$$

Since all $x_i$'s ensembles are canonical and independent, the average of $\exp(-\beta W_{\text{total}})$ is computed as

$$\left\langle e^{-\beta W_{\text{total}}} \right\rangle_{(x_0, x_1, ..., x_s)} = \frac{\int d\vec{r}^{3N-1} dx_0 e^{-\beta H_o(\vec{r}^{3N-1}, x_0)} e^{-\beta \frac{k}{2}(x_0 - \lambda_1)^2}}{Z(0)}$$

$$\times \frac{\prod_{i=1}^{s-1} \int d\vec{r}^{3N-1} dx_i e^{-\beta\left[H_o(\vec{r}^{3N-1}, x_i) + \frac{k}{2}(x_i - \lambda_i)^2\right]} e^{-\beta \frac{k}{2}\left[(x_i - \lambda_{i+1})^2 - (x_i - \lambda_i)^2\right]}}{Z(\lambda_1, k)...Z(\lambda_{s-1}, k)}$$

$$\times \frac{\int d\vec{r}^{3N-1} dx_s e^{-\beta\left[H_o(\vec{r}^{3N-1}, x_s) + \frac{k}{2}(x_s - \lambda_s)^2\right]} e^{+\beta \frac{k}{2}(x_s - \lambda_s)^2}}{Z(\lambda_s, k)}$$

$$= 1. \quad (B.2)$$

We first take the average of $\exp(-\beta W_{\text{total}})$ over $x_0$ and $x_1$ to obtain

$$\left\langle e^{-\beta W_{\text{total}}} \right\rangle_{x_0, x_s} = \frac{\int d\vec{r}^{3N-1} dx_0 e^{-\beta\left[H_o(\vec{r}^{3N-1}, x_0) + \frac{k}{2}(x_0 - \lambda_1)^2\right]}}{Z(0)}$$

$$\times \frac{\int d\vec{r}^{3N-1} dx_s e^{-\beta H_o(\vec{r}^{3N-1}, x_s)}}{Z(\lambda_s, k)} e^{-\beta W_{\text{mech}}}$$

$$= e^{\beta[F(\lambda_s, k) - F(\lambda_1, k)]} e^{-\beta W_{\text{mech}}}. \quad (B.3)$$

It is noted that based on Eq. (B.1) all possible values of $W_{\text{mech}}$ can be constructed from either force-versus-extension (FVE) curves or from the ensembles of $x_i$ with $i$ from 1 to $s-1$. Therefore, averaging the right hand side of Eq. (B3) over all possible FVE curves is equal to averaging the left hand side over the rest of $x_i$. Taking this observation into account and using the identity (B.2), we arrive at a formula identical to JE:

$$\Delta F^{\text{JE}}(\lambda_s, \lambda_1, k) = -\beta^{-1} \ln \left\langle e^{-\beta W_{\text{mech}}} \right\rangle_{\text{FVE}}. \quad (B.4)$$

Furthermore, $x_i$'s distributions can be approximated as $\exp[-\beta k(x_i - \langle x_i \rangle)^2/(2\gamma_i^2)]$, where $\langle x_i \rangle$ is the averaged position of the reaction coordinate at $i^{\text{th}}$ pulling step and $\gamma_i^2$ is equal to $k\sigma_i^2/k_B T$ with $\sigma_i$ equal to the standard deviation of $x_i$'s distribution. If all increments $\lambda_{i+1} - \lambda_i$ are the same as $\Delta\lambda = (\lambda_s - \lambda_1)/s$, we derive an analytical expression for the right hand side of Eq. (B.4):

$$\left\langle \exp[-\beta W_{\text{mech}}] \right\rangle_{\text{FVE}} = \frac{\int dW_{\text{mech}} \rho(W_{\text{mech}}) e^{-\beta W_{\text{mech}}}}{\int dW_{\text{mech}} \rho(W_{\text{mech}})} \quad (B.5)$$

$$\cong \prod_{i=1}^{s-1} \frac{\int dx_i \exp\left\{-\beta k\left[\frac{(x_i - \langle x_i \rangle)^2}{2\gamma_i^2} + \frac{\Delta\lambda(2\lambda_i + \Delta\lambda - 2x_i)}{2}\right]\right\}}{\int dx_i \exp[-\beta k((x_i - \langle x_i \rangle)^2/2\gamma_i^2)]}$$

$$= \prod_{i=1}^{s-1} \exp\left\{-\frac{\beta k \Delta\lambda^2}{2}[(1 - \gamma_i^2) + 2\frac{\lambda_i - \langle x_i \rangle}{\Delta\lambda}]\right\}.$$

Then the Gaussian-approximated free energy difference (FED) is given by

$$\Delta F^{G}(\lambda_s, \lambda_1, k) = \frac{k\Delta\lambda^2}{2}\sum_{i=1}^{s-1}(1-\gamma_i^2) + k\Delta\lambda\sum_{i=1}^{s-1}(\lambda_i - \langle x_i \rangle).$$

## APPENDIX C: WORK DISTRIBUTION CONSTRUCTION

Given $\lambda_1$, $\lambda_s$, $\Delta\lambda = (\lambda_s - \lambda_1)/s$ and a data set of $x_i$ we divide a sufficiently large interval, which covers all values of $x_i$, into $K$ bins with a bin width $\delta x$. The distributions $\rho_i(x_i)$ of $x_i$ are constructed by counting probability of $x_i$ falling into each bin. Similarly, we estimate a range having a bin width $\delta W$ and $2M+1$ bins that all mechanical work $W_{\text{mech}}$ ($W_1$, $W_2$... $W_{s-1}$) can fall into. For $i = 1$, $W_1$ is zero according to Eq. (9). For $i = 2$, we construct the distribution of $W_2$, $\Omega_2(W_2)$, as in the following pseudo-code:

**for** ($j = 1$; $j \leq K$; $j$++)
    **if** ($\rho_1(j) \neq 0$)   // $\rho_1(x_1)$ are non-zero in small regions around $\lambda_1$.
        $W_2 = k\Delta\lambda(2\lambda_1 - 2j\delta x + \Delta\lambda)/2$;
        $w = [W_2/\delta W]$; // a floor function transforms a real number into an integer.
        $\Omega_2(w) = \rho_1(j)$;    // all $\Omega_{i>2}$ are initialized to be zero.
    **endif**
**endfor**

For $i > 2$, the working distributions of these pulling steps are accumulated as the following:

**for** ($i = 3$; $i \leq s - 1$; $i$++)
    **for** ($j = 1$; $j \leq K$; $j$++)
        **if** ($\rho_{i-1}(j) \neq 0$)
            **for** ($w_1 = -M$; $w_1 \leq M$; $w_1$++)
                **if** ($\Omega_{i-1}(w_1) \neq 0$)
                      $W_i = w_1 \times \delta W + k\Delta\lambda(2\lambda_{i-1} - 2j\delta x + \Delta\lambda)/2$;
                      $w_2 = [W_i/\delta W]$;
                      $\Omega_i(w_2)$ += $\rho_{i-1}(j) \times \Omega_{i-1}(w_1)$;
                **endif**
            **endfor**
        **endif**
    **endfor**
**endfor**

We observe that the floor function produces an unwanted spike in work distributions at $W_i = 0.0$. We smooth work distributions at this value by assigning $\Omega_i(0) = [\Omega_i(-1) + \Omega_i(1)]/2$. From the distributions of work $\Omega_i$, it is straightforward to compute FEDs based on Eq. (12) or Eq. (B.5). The error analysis of these numerical calculations can be found elsewhere [18, 38]. The variance of FEDs can be estimated by $\sigma_W^2/Q + \beta^2\sigma_W^4/2(Q-1)$, where $\sigma_W$ is the standard deviation of work distributions $\Omega_i(W_i)$ and $Q$ is a number of bins which have non-zero $\Omega_i(W_i)$. For example, at pulling step $i = 35$, $\sigma_W$ is about 10 kcal/mol with $Q \sim 20000$, then the variance is about 0.7 kcal/mol at temperature $T = 300$ K. Thus the standard deviation of the corresponding FED is about 0.8 kcal/mol. This deviation is about two times smaller than half the difference (~ 2.0 kcal/mol) between $\Delta F^{JE}(\lambda_s, \lambda_1, k)$ and $\Delta F_{\text{fluct}}(\lambda_s, \lambda_1, k)$ in Sections IV B and D. Larger $Q$ value (larger $K$ and $M$) does not make the difference smaller, even though it reduces the deviation. Therefore, we choose $[\Delta F^{JE}(\lambda_s, \lambda_1, k) - \Delta F_{\text{fluct}}(\lambda_s, \lambda_1, k)]/2$ as the major uncertainty of our method. We find that $\delta x = 0.001$ Å and $\delta W = 0.01$ kcal/mol give reasonable estimates of work distributions and their FEDs investigated in the test case.

*Electronic address: nvan@usc.edu
[1] C. Jarzynski, Physical Review Letters **78**, 2690 (1997).
[2] J. T. Liphardt *et al.*, Biophysical Journal **82**, 193A (2002).
[3] D. Collin *et al.*, Nature **437**, 231 (2005).
[4] F. Douarche, and et al., EPL (Europhysics Letters) **70**, 593 (2005).
[5] D. A. Hendrix, and C. Jarzynski, The Journal of Chemical Physics **114**, 5974 (2001).
[6] C. Jarzynski, Comptes Rendus Physique **8**, 495.
[7] C. Jarzynski, Physical Review E **56**, 5018 (1997).
[8] C. Jarzynski, Proc Natl Acad Sci U S A **98**, 3636 (2001).
[9] C. Jarzynskia, The European Physical Journal B - Condensed Matter and Complex Systems **64**, 331 (2008).
[10] G. Crooks, Journal of Statistical Physics **90**, 1481 (1998).
[11] G. E. Crooks, Physical Review E **60**, 2721 (1999).
[12] G. Hummer, and A. Szabo, Proceedings of the National Academy of Sciences of the United States of America **98**, 3658 (2001).
[13] G. Hummer, and A. Szabo, Biophysical Journal **85**, 5 (2003).
[14] G. Hummer, and A. Szabo, Abstracts of Papers of the American Chemical Society **227**, U897 (2004).
[15] G. Hummer, and A. Szabo, Accounts of Chemical Research **38**, 504 (2005).
[16] G. Hummer, and A. Szabo, Proceedings of the National Academy of Sciences of the United States of America **107**, 21441 (2010).
[17] S. Park *et al.*, Journal of Chemical Physics **119**, 3559 (2003).
[18] D. Rodriguez-Gomez, E. Darve, and A. Pohorille, Journal of Chemical Physics **120**, 3563 (2004).
[19] H. Oberhofer, C. Dellago, and P. L. Geissler, Journal of Physical Chemistry B **109**, 6902 (2005).
[20] D. K. West, P. D. Olmsted, and E. Paci, The Journal of Chemical Physics **125**, 204910 (2006).
[21] S. Park, and K. Schulten, The Journal of Chemical Physics **120**, 5946 (2004).
[22] D. D. L. Minh, and A. B. Adib, Physical Review Letters **100** (2008).

[23] B. Isralewitz, M. Gao, and K. Schulten, Current Opinion in Structural Biology **11**, 224 (2001).
[24] G. Hummer, Journal of Chemical Physics **114**, 7330 (2001).
[25] E. Darve, and A. Pohorille, Journal of Chemical Physics **115**, 9169 (2001).
[26] T. Bastug *et al.*, The Journal of Chemical Physics **128**, 155104 (2008).
[27] T. Bastug, and S. Kuyucak, Chemical Physics Letters **436**, 383 (2007).
[28] G. M. Torrie, and J. P. Valleau, Journal of Computational Physics **23**, 187 (1977).
[29] S. Kumar *et al.*, Journal of Computational Chemistry **13**, 1011 (1992).
[30] M. Souaille, and B. Roux, Computer Physics Communications **135**, 40 (2001).
[31] P. Cieplak, and P. A. Kollman, Journal of Computer-Aided Molecular Design **7**, 291 (1993).
[32] G. Ozer *et al.*, Journal of Chemical Theory and Computation **6**, 3026 (2010).
[33] R. W. Zwanzig, The Journal of Chemical Physics **22**, 1420 (1954).
[34] G. Schay, *Introduction to Probability with Statistical Applications* (Birkhauser Boston, Boston, 2007), p. 313.
[35] M. P. Allen, and D. J. Tildesley, *Computer Simulation of Liquids* (Oxford University Press, 2003), p. 213.
[36] L. Kale *et al.*, Journal of Computational Physics **151**, 283 (1999).
[37] A. D. MacKerell *et al.*, Journal of Physical Chemistry B **102**, 3586 (1998).
[38] D. M. Zuckerman, and T. B. Woolf, Physical Review Letters **89** (2002).
[39] If the PPP is used in the case of k = 1.0 kcal/mol/Å2, one might have to use larger τ to have good statistics of the reaction coordinate at large λ than τ at small λ.
[40] The range of values of work is enlarged to 450 kcal/mol to cover work distributions for either large k or small τ.
[41] J. M. G. Vilar, and J. M. Rubi, Physical Review Letters **100** (2008).
[42] L. Peliti, Physical Review Letters **101** (2008).
[43] J. Horowitz, and C. Jarzynski, Physical Review Letters **101** (2008).
[44] E. G. D. Cohen, and D. Mauzerall, Journal of Statistical Mechanics-Theory and Experiment (2004).
[45] D. A. Kofke, Molecular Physics **104**, 3701 (2006).
[46] N. D. Lu, D. A. Kofke, and T. B. Woolf, Journal of Computational Chemistry **25**, 28 (2004).
[47] G. E. Crooks, Physical Review E **61**, 2361 (2000).
[48] P. Kollman, Chemical Reviews **93**, 2395 (1993).
[49] S. Miyamoto, and P. A. Kollman, Proteins-Structure Function and Bioinformatics **16**, 226 (1993).
[50] T. P. Straatsma, H. J. C. Berendsen, and A. J. Stam, Molecular Physics **57**, 89 (1986).